\documentclass[]{emulateapj}

\usepackage{natbib}
\usepackage{hyperref}

\shorttitle{DIRBE Comet Trails}
\shortauthors{Arendt}

\begin{document}

\title{DIRBE Comet Trails}

\author{Richard G. Arendt}
\affil{CREST/UMBC, Code 665, NASA/GSFC, Greenbelt, MD 20771; Richard.G.Arendt@nasa.gov}

\begin{abstract}
Re-examination of the {\it COBE} DIRBE data reveals the 
thermal emission of several comet dust trails. 
The dust trails of 1P/Halley, 169P/NEAT, and 3200 Phaethon have not been previously reported.
The known trails of 2P/Encke, and 73P/Schwassmann-Wachmann 3 are also seen.
The dust trails have 12 and 25 $\mu$m surface brightnesses of 
$<0.1$ and $<0.15$ MJy sr$^{-1}$, respectively, which is $<1\%$ of the zodiacal light intensity.
The trails are very difficult to see in any single daily image of the sky, but 
are evident as rapidly moving linear features in movies of the DIRBE data.
Some trails are clearest when crossing through the orbital plane of the 
parent comet, but others are best seen at high ecliptic latitudes as the Earth passes over 
or under the dust trail. All these comets have known associations with meteor showers.
This re-examination also reveals one additional comet and 13 additional asteroids 
that had not previously been recognized in the DIRBE data.

\end{abstract}


\section{Introduction}

The first associations between annual meteor showers and periodic comets 
were made in the 1860s by \cite{Schiaparelli:1867} and others. Accounts of the history of these early 
developments are presented by \cite{Kirkwood:1873}, \cite{Littmann:1998} and \cite{Jenniskens:2006}.
This led to the recognition that the meteors observed in annual showers are the debris shed 
as comets break up and disintegrate over time with repeated passages near the sun.
Over the next century, the number of associations between meteor showers and comets grew
with accumulation of additional data on both comets and meteors, and with advances in 
the computational capabilities in determining the perturbations and evolution of orbits
of short period comets and associated meteoroid 
streams \citep[and references therein]{Jenniskens:2006}.

A fundamentally new phase of this research began in 1983 with the launch of the
{\it Infrared Astronomical Satellite} \citep[{\it IRAS},][]{Neugebauer:1984}. {\it IRAS}
performed the first full-sky survey at mid-IR wavelengths of 12, 25, 60, and 100 $\mu$m. 
At these wavelengths, particularly 12 and 25 $\mu$m, the observations are very
often dominated by the thermal emission of interplanetary dust, or zodiacal light.
The orbits of interplanetary dust grains are not stable, with the grains 
slowly spiraling inward under the influence of Poynting-Robertson and solar wind
drag, and radiation pressure \citep{Burns:1979, Ipatov:2008}. 
Therefore it had been recognized that 
the interplanetary dust needs to be continually replenished from new sources of
dust. {\it IRAS} provided clear evidence for replenishment by main belt asteroids 
with the discovery of ``bands'' at low ecliptic latitudes
\citep{Low:1984, Sykes:1988, Nesvorny:2006}. {\it IRAS} also provided
evidence of replenishment by comets with ``the discovery of dust trails in the 
orbits of periodic comets'' \citep{Sykes:1986}.  With {\it IRAS} it became 
possible to {\it see} the dust trails of debris that is shed from comets. These
trails provide a means of detecting and quantifying (nascent) meteoroid streams
even in cases where the streams do not intersect or approach Earth's orbit \citep{Sykes:1992}.
Later space-based IR observatories, {\it ISO} and {\it Spitzer}, have discovered additional 
dust trails and have provided observations with greatly improved spectral and spatial
resolution. Recently, dust trails have also been
detected at optical wavelengths with ground-based instruments \citep{Ishiguro:2009}.Earth
These studies provide detailed characterization of dust grain
size and composition, and on the total masses of the 
trails \citep[e.g.][]{Sykes:1990, Reach:2007, Reach:2009, Vaubaillon:2010}.

In 1989-90 the Diffuse Infrared Background Experiment (DIRBE) on the {\it Cosmic
Background Explorer (COBE)} satellite, like {\it IRAS}, also performed a 
full-sky survey at mid-IR wavelengths \citep{Hauser:1998}. DIRBE was designed for measurement of the
cosmic IR background, and to provide complementary observations to the 
microwave background experiments on {\it COBE}. Therefore it was designed 
with the capability of making absolutely calibrated
brightness measurements over a wide range of solar elongations, but
with low angular resolution. Several of the brightest comets in the inner
solar system were detected by DIRBE, but no dust trails of these or
other comets were identified \citep{Lisse:1998}.

Despite this negative result, the recent dusty demise of comet ISON
\citep{Knight:2014}
has inspired a reinvestigation of the DIRBE data to see if there
might be a detectable dust trails associated with other sun-grazing comets. 
In particular, the Kreutz family sungrazers \citep{Marsden:1989, Marsden:2005} seemed like a likely 
candidate for having an old dust trail formed from multiple past break ups, 
as well as a new dust trail forming from the disintegration of the hundreds 
of small comets witnessed by the {\it Solar \& Heliospheric Observatory (SOHO)} 
and other solar coronagraphic satellites over the past decades \citep{Knight:2010}.

The results of an initial reinvestigation of the DIRBE data are reported here.
Section 2 of this paper provides a brief overview of the DIRBE data, and describes 
the specific data reduction steps that were taken that greatly enhance the 
detection of faint moving objects and structures in the solar system.
Section 3 summarizes asteroids and comets that are newly revealed in the DIRBE data.
Section 4 provides a guide to the comet dust trails that are now evident in the 
DIRBE data. A key aspect of recognizing the dust trails is the inspection of 
animations of daily images of the IR sky brightness after removal of
bright large-scale zodiacal light. The discussion in Section 5 includes the basic characterization
of the width and brightness of the dust trails, and limitations on DIRBE's
ability to detect trails. It also describes associations between the trails and
meteor showers, and discusses possibilities for further improvements in 
reducing the DIRBE data for the purpose of detecting dust trails. 
The paper is summarized in Section 6.

\section{DIRBE Data}

The DIRBE instrument was built to measure the absolute brightness of the entire sky at 
$\sim0.7\arcdeg$ resolution in 10 broad bands at $\lambda =$ 1.25, 2.2, 3.5, 4.9, 12, 25, 60, 100,
140, and 240 $\mu$m. The DIRBE beam was set at a $30\arcdeg$ angle from the spacecraft 
spin axis, and thus traced a $60\arcdeg$ diameter circle on the sky with each 0.8 rpm 
rotation of the spacecraft. The {\it COBE} spin axis was kept pointing near the local zenith 
(at a solar elongation of $\sim94\arcdeg$) as the spacecraft orbited in a polar orbit with a period of 
103 min. In a single orbit the DIRBE beam would therefore trace a cycloidal path over a 
viewing swath covering solar elongations $64\arcdeg \lesssim \epsilon \lesssim 124\arcdeg$.
The coverage of this swath is sparse for a single orbit, but is fairly complete  
after a full day (though still shallow). Averaging data on a weekly timescale produces high quality images of the 
viewing swath, but can hide details of the changing zodiacal light and moving solar system objects.
The accumulation of sky coverage by DIRBE is nicely illustrated by Figure 1 of \cite{Kelsall:1998}. Nominal cryogenic operations 
of all the DIRBE bands lasted for 285 days ($\sim3/4$ yr), which was sufficient to obtain complete
coverage of the full sky, but not with the uniformity that a full year of operation would have provided.
Details of the DIRBE instrument can be found in \cite{Silverberg:1993}, \cite{Hauser:1998}, and the 
DIRBE Explanatory Supplement\footnote{\url{http://lambda.gsfc.nasa.gov/product/cobe/dirbe_exsup.cfm}}. 
Information on the {\it COBE} spacecraft and mission is presented
by \cite{Boggess:1992}. 

The reprocessing of the DIRBE data for the present analysis 
began with the Calibrated Individual Observations. For the appropriate
time and location of each observation, the \cite{Kelsall:1998} model of the zodiacal light was calculated
for all DIRBE bands.
The observations of each day were averaged into separate sky maps with and without subtraction
of the zodiacal light. Observations within $10\arcdeg$ of the moon were excluded from the averages.
These zodiacal light subtracted images are not well suited for detection of faint structure and moving
objects because they still contain the Galactic background and residuals of the zodiacal light subtraction.
At $\lambda \geq 12$ $\mu$m the Galactic background is dominated by thermal emission of dust in 
the ISM, while at $\lambda \leq 4.9$ $\mu$m stellar sources dominate.
Most of the residual zodiacal light emission retains a relatively fixed pattern with respect to elongation 
and ecliptic latitude, modulated by a slower evolution of this pattern during the mission. 

\begin{figure*}[ht] 
\centering
\includegraphics[height=5.6in]{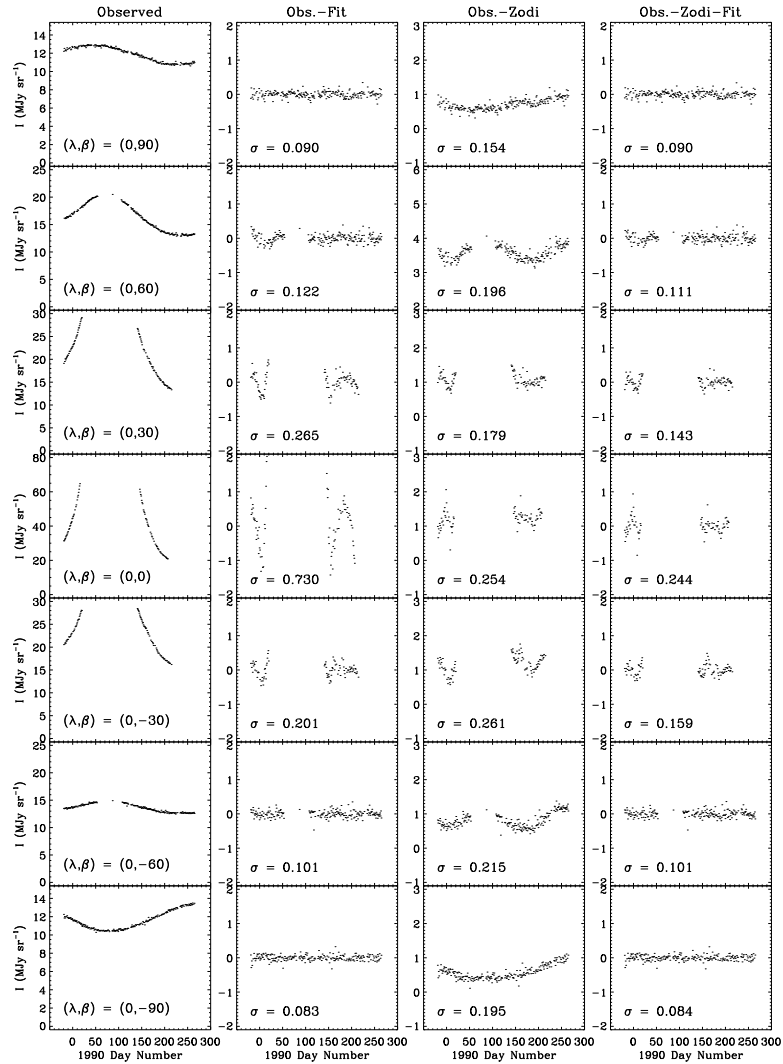} 
\caption{\small The temporal variation of the 12 $\mu$m brightness at representative
ecliptic latitudes. The left column shows the observed sky brightness. The second
column shows the brightness after subtraction of a simple empirical fit
to the variation at each pixel (Eq. \ref{eq:fit}). The third column shows the residual variation 
after the \cite{Kelsall:1998} zodiacal light model is subtracted from the observations.
The last column shows the results after application of Eq. \ref{eq:fit} to the residual
variation after subtraction of the zodiacal light model. The ecliptic coordinates
of the fields are noted in the first column. The standard deviation of the residual 
variations (in MJy sr$^{-1}$) are listed in the other columns.
Subtraction of both the zodiacal light model and the low frequency fit
to the residuals is needed to minimize variations at all latitudes.
\label{fig:plot-zodi-b5}}
\end{figure*}

\begin{figure*}[ht] 
\centering
\includegraphics[height=5.6in]{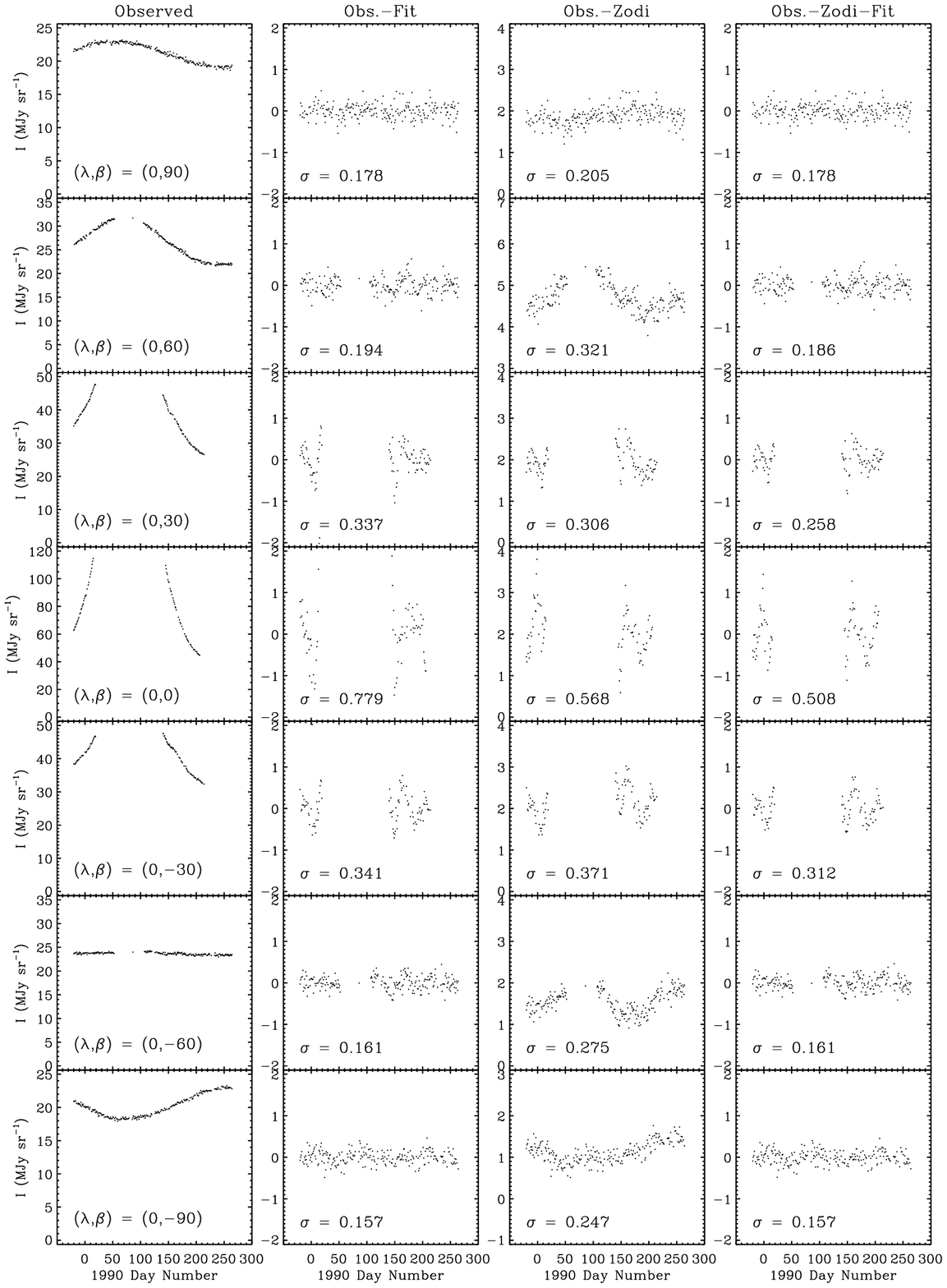} 
\caption{Temporal variations and residual emission at 25 $\mu$m, as 
described in Fig. \ref{fig:plot-zodi-b5} \label{fig:plot-zodi-b6}}
\end{figure*}

To mitigate these variations, the time series of observations were filtered to remove the lowest
frequency components. The time series at each pixel was fit by:
\begin{eqnarray}
I_\nu & = & A_0 + A_1 \cos{(2\pi t)} + B_1 \sin{(2\pi t)} + \nonumber \\ 
    &   & A_2 \cos{(2\ 2\pi t)} + B_2 \sin{(2\ 2\pi t)}
\label{eq:fit}
\end{eqnarray}
where time $t$ is measured in years.
The derived coefficients correspond to the real and imaginary amplitudes of the first 3 terms in the 
Fourier decomposition of the variation at each pixel. Tests were done subtracting additional higher frequency 
components, or separately fitting observations in the leading or trailing halves of the orbit, or
with different functional (e.g. polynomial) forms. However, these more complex versions were
not used because further improvements were small, additional artifacts were introduced, and/or
the additional degrees of freedom in the fit began to remove real features of interest.
Another alternate test of this technique was to apply the fit to the data without
prior subtraction of the zodiacal light model. This works fairly well as an ad hoc zodiacal 
light subtraction at high latitudes, but still leaves substantial residuals at low latitudes
where the zodiacal light is brighter and more structured. 
Figures \ref{fig:plot-zodi-b5} and \ref{fig:plot-zodi-b6} illustrate the effectiveness 
of the removal of the temporal variations at 12 and 25 $\mu$m for representative pixels 
at various ecliptic latitudes. The original variations are shown along with the different versions 
of zodiacal light and fit subtracted results.

Subtraction of the zodiacal light and the fit from each pixel makes significant improvement 
in flatness of the residual images. The constant term in the fit is very effective at removing the 
diffuse Galactic emission. However, bright point 
sources still leave residuals because the square shape and sharp edges of the DIRBE beam,
combined with the variable scan direction when crossing a given source, induce an 
irregular high-frequency variability in pixels at the edges of the sources. This is the biggest 
limiting factor in the short wavelength images, $\lambda \leq 4.9$ $\mu$m, which will not
be discussed in further detail here. 
At wavelengths $\lambda \geq 60$ $\mu$m residual calibration defects cause temporal drifts to 
map into large angular scale structure matching the scan pattern. This limits the usefulness of the 
these data at the present time. Thus the remaining analysis performed here utilizes the 
12 and 25 $\mu$m results. These wavelengths are where the emission of interplanetary dust 
is brightest and has the highest contrast with respect to the Galactic background.

\begin{figure*}[ht] 
\centering
\includegraphics[width=4.5in]{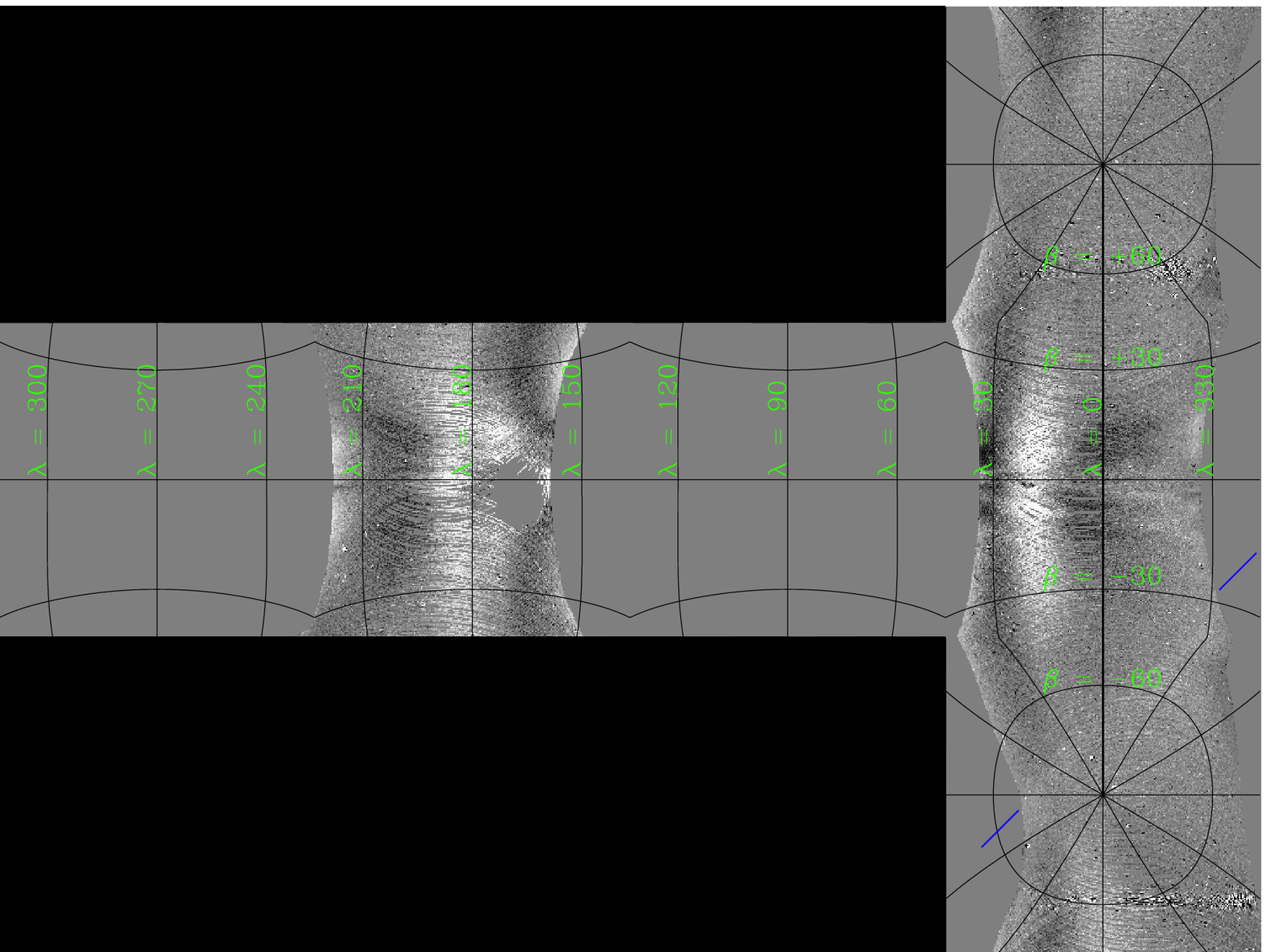} 
\caption{Still frame (Day 90179) from 12 $\mu$m residual movie in sky cube format,
with an ecliptic coordinate grid superimposed. The trail of 2P/Encke is faintly visible 
across the SEP (between the blue tick marks). This image is scaled linearly from 
$-0.5$ to 0.5 MJy sr$^{-1}$. The 12 $\mu$m movies without and with annotation 
are shown in (a) and (b). The corresponding 25 $\mu$m movies are in (c) and (d).
\label{fig:cube}}
\end{figure*}

At each wavelength, the sets of daily images were assembled into movies of the 
sky over the 285 days of the cryogenic mission. Movies in the native COBE sky cube 
format\footnote{\url{http://lambda.gsfc.nasa.gov/product/cobe/skymap_info_new.cfm}} \citep{Greisen:2006}
offer the most accurate representation of the data (Figure \ref{fig:cube}a-d), and are good for examining features that
are at either moderately low or moderately high latitudes ($|\beta| \lesssim30\arcdeg$ or  
$|\beta| \gtrsim 60\arcdeg$). The initial identifications of most objects and trails were made
in these movies, as described in the following sections. The still frame in Figure \ref{fig:cube} shows 
the ecliptic coordinates superimposed on a single daily image in the sky cube projection.

\begin{figure*}[ht] 
\centering
\includegraphics[width=4.5in]{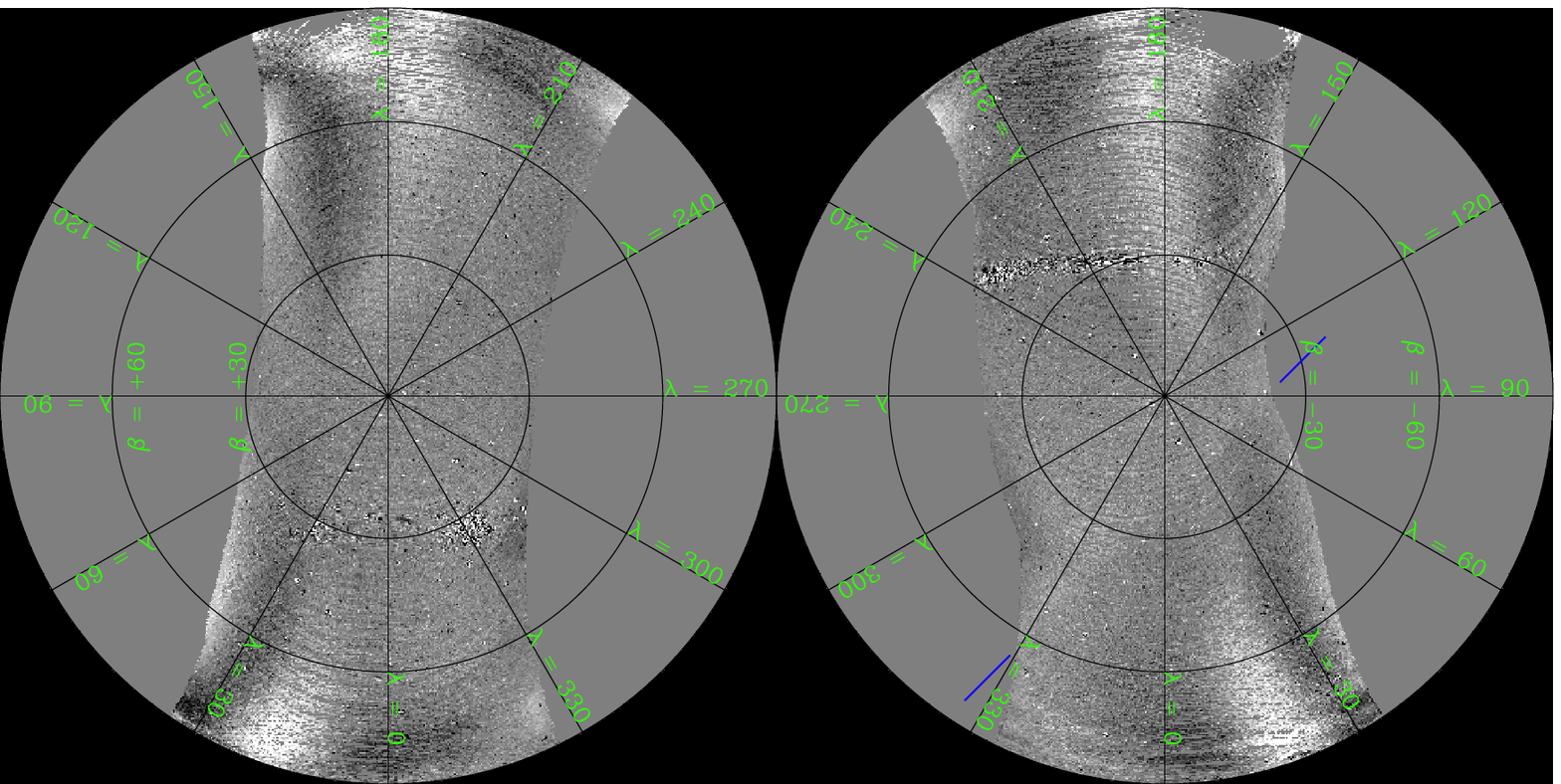} 
\caption{Still frame (Day 90179) from 12 $\mu$m residual movie in polar format, 
with an ecliptic coordinate grid superimposed. The trail of 2P/Encke is faintly visible 
across the SEP (between the blue tick marks). This image is scaled linearly from 
$-0.5$ to 0.5 MJy sr$^{-1}$. The 12 and 25 $\mu$m movies are shown in (a) and (b).
Additional ``cutouts'' from the 12 $\mu$m movie loop back-and-forth over short intervals to highlight 
the motions of the trails of (c) 1P/Halley, (d) 2P/Encke, (e) 73P/SW3, (f) and (g) 
169P/NEAT, (h) and (i) 3200 Phaethon.
\label{fig:pole}}
\end{figure*}

Movies of the daily images transformed into polar projections also prove
useful (Figure \ref{fig:pole}a-b). These are especially good for tracing the full extent of trails when they span a large range
in latitude, thus being interrupted at the boundaries between the equatorial and polar
cube faces in the sky cube projection. The weakness of the polar projected movies is that 
they do no clearly show features at very low ecliptic latitudes.
The still frame in Figure \ref{fig:pole} shows 
the ecliptic coordinates superimposed on a single daily image in the polar projection.

Each of the movies is produced in a bare version and an annotated version.
The annotated versions include labels tracking the positions of asteroids and comets 
that can be seen at some point in the data. Labels similarly mark the locations of the 
bodies with associated dust trails, although these bodies are not detected except for 
73P / Schwassmann-Wachmann 3. The projected orbits of the bodies with associated 
dust trails are marked with $+$ symbols at 1-day increments along the orbits. The 
perihelions are marked with larger solid dots. Points at mean anomalies of $90\arcdeg$, 
$180\arcdeg$ (aphelion) and $270\arcdeg$ are marked as ``1'', ``2'', ``3''.

\section{Moving Objects}

Close inspection of the 12 and 25 $\mu$m DIRBE movies reveals more than 20 moving solar system objects. 
Mars, Jupiter, and Saturn are obvious bright sources. Uranus is present in the DIRBE data, but it 
is not seen as a moving object in the movies. The four comets studied by \cite{Lisse:1998} are easily seen,
and an additional comet, C/1989 T1 (Helin-Roman-Alu), can be seen passing across the north ecliptic pole (NEP)
from the start of the mission 
until day 90066\footnote{DIRBE day numbers are formatted as a 2-digit year + a 3-digit day of year; 90066 = 
the 66th day of 1990 = 1990 Mar 07}. 
The remaining moving objects are asteroids. Most are only visible for a fraction of the time that
they are within the DIRBE viewing swath. 

These solar system objects are listed in Table \ref{tab:sso} and are their locations are noted in the 
annotated DIRBE movies. No flux densities were extracted for these objects because more accurate
photometry can be obtained from older {\it IRAS} or newer {\it WISE} data, and because
photometry is better done in the time domain \citep[e.g.][]{Lisse:1998,Smith:2004}, than in
the daily averaged images constructed here.

\begin{deluxetable*}{llllll}
\tablewidth{0pt} 
\tabletypesize{\footnotesize}
\tablecaption{Moving Solar System Objects Detected by DIRBE}
 \tablehead{
\colhead{Planet} & 
\colhead{Reference} &
\colhead{Asteroid} & 
\colhead{Reference} &
\colhead{Comet} & 
\colhead{Reference}
} 
\startdata
Mars    & DIRBE\tablenotemark{1} & 1 Ceres      & DIRBE\tablenotemark{1} & 73P/Schwassmann-Wachmann 3                  & \cite{Lisse:1998} \\
Jupiter & DIRBE\tablenotemark{1} & 2 Pallas     & DIRBE\tablenotemark{1} & C/1989 Q1 (Okazaki-Levy-Rudenko)              & \cite{Lisse:1998} \\
Saturn  & DIRBE\tablenotemark{1} & 4 Vesta      & DIRBE\tablenotemark{1} & C/1989 T1 (Helin-Roman-Alu)       &new \\
        &       & 15 Eunomia    & new   & C/1989 X1 (Austin)  & \cite{Lisse:1998} \\
        &       & 31 Euphrosyne     & new   & C/1990 K1 (Levy)     &  \cite{Lisse:1998} \\
        &       & 41 Daphne     & new \\
        &       & 42 Isis       & new \\
        &       & 85 Io         & new \\
        &       & 185 Eunike & new \\
        &       & 194 Prokne     & new \\
        &       & 372 Palma      & new \\
        &       & 405 Thia       & new \\
        &       & 511 Davida     & new \\
        &       & 704 Interamnia & new \\
        &       & 747 Winchester & new \\
        &       & 1021 Flammario  & new 
\enddata
\tablenotetext{1}{DIRBE Solar System Objects Data: \url{http://lambda.gsfc.nasa.gov/product/cobe/dirbe_products.cfm}}
\label{tab:sso}
\end{deluxetable*}

The orbital data for all moving objects in the study were obtained from the 
JPL Horizons system \citep{Giorgini:1996}. Ephemerides were generated for tracking the locations
of specific moving objects during the DIRBE mission, 
and orbital elements were used for plotting orbits projected onto DIRBE data.

\section{Comet Trails}

\begin{figure}[ht] 
\centering
\includegraphics[width=3.in]{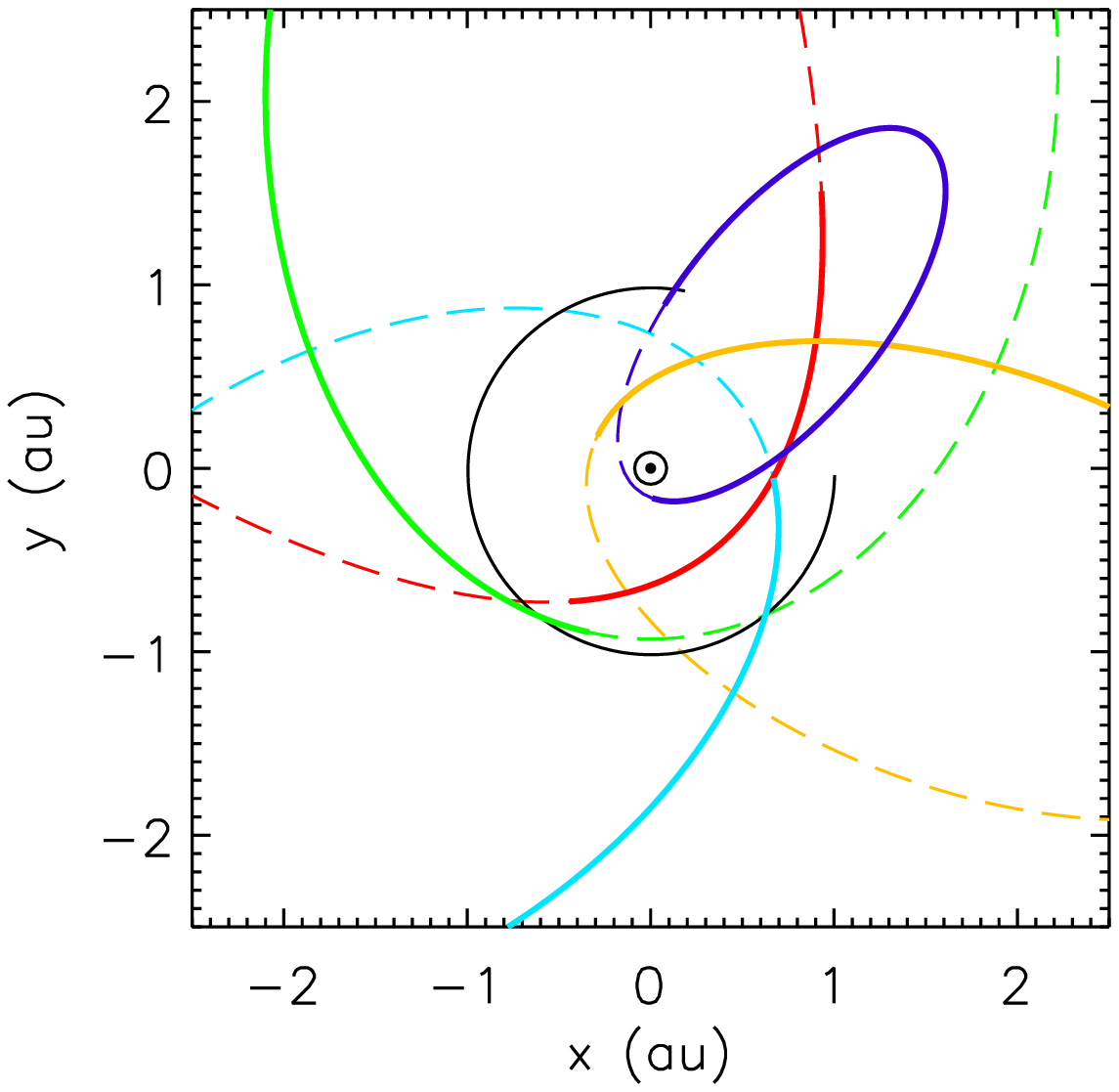} 
\caption{Comet orbits projected on the ecliptic plane. 1P/Halley = red, 2P/Encke = Orange, 73P/SW3 = green, 
169P/NEAT = cyan, 3200 Phaethon = violet. Dashed lines indicate portions of orbits south of the ecliptic plane.
Earth's orbit (black) is only shown for duration of the {\it COBE} cryogenic mission from 89345 - 90264.
The dust trails are usually most visible at these orbital crossings.}
\label{fig:orbits}
\end{figure}

Trails were expected to be most prominent when viewed from the plane of the comet orbit 
(when the Earth passes the line of nodes). However, trails were also found to be visible at high 
latitudes as the Earth crosses above of below the comet orbit at $\sim1$ au (see Fig. \ref{fig:orbits}).
The following text (summarized in Tables  \ref{tab:nodes} and \ref{tab:crossing})
describes when and where the dust trails reported here can be seen in 
the DIRBE data.

\begin{deluxetable*}{llllll}
\tabletypesize{\scriptsize}
\tablewidth{0pt}
\tablecaption{DIRBE Comet Trails at Lines of Nodes}
\tablehead{
\colhead{} & \colhead{1P/Halley} & \colhead{2P/Encke} & \colhead{73P/SW3} & \colhead{169P/NEAT} & \colhead{3200 Phaethon}
}
\startdata
Ascending Node Date   &  \nodata & {\bf 90240} & \nodata & 90076 & 90168 \\
Ascending $\Delta r$ (au)   & \nodata & {\bf 2.94} & \nodata & 1.71 & -0.87 \\
Dates Detected                & \nodata & {\bf 90233-90257?} & \nodata & none & $\epsilon < 64\arcdeg$ \\
Mean Anomaly Range\tablenotemark{a} ($\arcdeg$) & \nodata & {\bf 302.6 -- 332.8} & \nodata & \nodata & \nodata\\
Meteor Shower & \nodata & \nodata & \nodata & \nodata & \nodata \\
\hline
Descending Node Date & {\bf 90140} & 90054 & {\bf 90152} & 90262 & {\bf 89351}  \\
Descending $\Delta r$ (au)   & {\bf -0.16} & -0.65 & {\bf -0.06} & -0.33 & {\bf -0.09} \\
Trail Detected                & {\bf 90126-90142} & $\epsilon < 64\arcdeg$ & {\bf 90145-90154?} & none ($\epsilon < 66\arcdeg$) & {\bf 89349-89357} \\
Mean Anomaly Range\tablenotemark{a} ($\arcdeg$) & {\bf 0.48 -- 0.97} & \nodata & {\bf 358.7 -- 360.0?} & \nodata & {\bf 328.4 -- 333.2 \& 90.0-200.0}\\
Meteor Shower & \nodata & \nodata & $\tau$ Herculids & \nodata & \nodata
\enddata
\tablenotetext{a}{Minimum and maximum mean anomaly (measured from perihelion) at which a dust trail is visible during the dates listed.}
\label{tab:nodes}
\end{deluxetable*}

\begin{deluxetable*}{llllll}
\tabletypesize{\scriptsize}
\tablewidth{0pt}
\tablecaption{DIRBE Comet Trails at Projected Earth Orbit Crossing (NEP or SEP)}
\tablehead{
\colhead{} & \colhead{1P/Halley} & \colhead{2P/Encke} & \colhead{73P/SW3} & \colhead{169P/NEAT} & \colhead{3200 Phaethon}
}
\startdata
Inbound Date                  &  \nodata & \nodata & {\bf 90134} & {\bf 90212} & 89348 \\
Inbound $\Delta z$ (au) & \nodata   & \nodata & {\bf +0.06} & {\bf +0.15} & +0.02 \\
Inbound Detected & \nodata & \nodata & {\bf 90084-90136} & {\bf 90206-90224} & none? \\
Mean Anomaly Range\tablenotemark{a} ($\arcdeg$) & \nodata & \nodata & {\bf 348.6 -- 355.4} & {\bf 335.4 -- 350.2} & \nodata \\
Meteor Shower & \nodata & \nodata & \nodata & $\alpha$ Capricornids & Geminids \\
\hline
Outbound Date                  & {\bf 90127} & {\bf 90183} & 90212 & {\bf 90019} & \nodata \\
Outbound $\Delta z$ (au) & {\bf -0.07}   & -{\bf 0.17} & -0.17 & {\bf -0.17} & \nodata \\
Outbound Detected & {\bf 90126-90142} & {\bf 90143-90211} & none & {\bf 90010-90040} & {\bf 90243-90264} \\
Mean Anomaly Range\tablenotemark{a} ($\arcdeg$) & {\bf 0.48 -- 0.97} & {\bf 11.1 -- 41.9} & \nodata & {\bf 8.4 -- 21.8} & {\bf 23.4 -- 61.8}\\
Meteor Shower & $\eta$ Aquariids & Daytime $\zeta$ Perseids & \nodata & Daytime $\chi$ Capricornids & \nodata
\enddata
\tablenotetext{a}{Minimum and maximum mean anomaly (measured from perihelion) at which a dust trail is visible during the dates listed.}
\label{tab:crossing}
\end{deluxetable*}

\subsection{1P/Halley}
This dust trail is brightest, and most easily identified, in the direction of the tangent point 
along the orbit when passing the descending node of the orbit near day 90140. 
Prior to that, from as early as day 90126, the trail can be seen very faintly sweeping 
across the south ecliptic pole (SEP) as the Earth crosses over the trail. 
(See the cutout movie Figure \ref{fig:pole}c.)

The opposite node and crossing were not observed by DIRBE. At the ascending node, 
the Earth is inside Halley's orbit, thus the orbit is projected as a great circle and there are no 
tangent points. When crossing under Halley's trail, the distance is greater than when crossing 
over the trail. Thus the trail is not expected to appear as bright at either of these times. 

\subsection{2P/Encke}
The dust trail of 2P/Encke sweeping across the SEP from day 90143 to 90211 is the most obvious 
of the dust trails. It is also very asymmetric. It appears to be sharply bounded on the 
inside, at smaller heliocentric distances, but outside of the orbit it appears to fade slow into
the background over a distance of $>10\arcdeg$. 
(See the cutout movie Figure \ref{fig:pole}d.)
The crossing under 2P/Encke's orbit was not 
observed by DIRBE, but is likely to be similarly prominent. 

When passing through the descending node of the orbit, the entire orbit lies at low 
elongations ($\epsilon < 64\arcdeg$) and thus could not be observed by DIRBE. 
When passing through the ascending node near day 90240, the trail 
may be visible for a few days when passing through the plane, but only on the 
northern side of the orbit. Confusion with the Galactic plane and residual zodiacal light 
is lower in the north than in the south part of the orbit, but the proximity to 2P/Encke is likely 
to be more important in making the northern portion more visible.

\subsection{73P/Schwassmann-Wachmann 3 (SW3)}
This trail is first evident following day 90084, along the orbit immediately behind 73P/SW3. 
It is most visible as Earth crosses under the orbit (the trail sweeps across the NEP) 
near day 90134. 
(See the cutout movie Figure \ref{fig:pole}e.)
The trail fades but it remains marginally visible up until 
day 90152 when the Earth passes the descending node.

Passage through the ascending node was not observed by DIRBE, but would be projected as 
a great circle and distant, and thus would likely be faint. The Earth crosses over the orbit of 73P/SW3 at
day 90212, however the trail is not visually evident at the this time. At this point the trail 
is $\sim3$ times more distant than when the Earth crosses under the trail.
The trail might be detected here by averaging along its expected location, but 
the significance is low.

\subsection{169P/NEAT}
The trail of 169P/NEAT is evident both when Earth crosses above and below the trail. 
It is fainter than the trail of 2P/Encke, but it also seems to have a similar asymmetric profile.
(See the cutout movies Figures \ref{fig:pole}f-g.)

The trail is not evident when projected as a great circle as Earth passes the ascending node 
on day 90076. When passing the descending node at day 90262, nearly the entire orbit lies
at solar elongation, $\epsilon < 64\arcdeg$. One of the tangent points is barely within 
the viewing swath, but is too confused by residual zodiacal light artifacts to be detected.

\subsection{3200 Phaethon}
The sweep of Phaethon's trail across the NEP as the Earth passes underneath is not
evident, but this is probably because the apparent daily motion is so fast that the trail gets 
averaged away and/or smeared when constructing daily images. Shortly afterwards 
the trail brightens dramatically as the Earth passes through the orbital plane 
at the descending node on day 89351. 
(See the cutout movies Figures \ref{fig:pole}h.)
At this time  it appears that both the near and far sides of the orbit are
visible, especially at 25 $\mu$m (Fig. \ref{fig:pole}b).
The entire orbit lies at low elongation ($\epsilon <64\arcdeg$)
when passing through the ascending node. DIRBE did not observe when the 
Earth crosses back under Phaethon's trail, but prior to this the trail is 
faintly detected sweeping northward from day 90234 until the end of the 
cryogenic mission on 90264.
(See the cutout movies Figures \ref{fig:pole}i.)

\section{Discussion}

\subsection{Characterization of the Dust Trails} 

\begin{figure}[ht] 
\centering
\includegraphics[width=3.in]{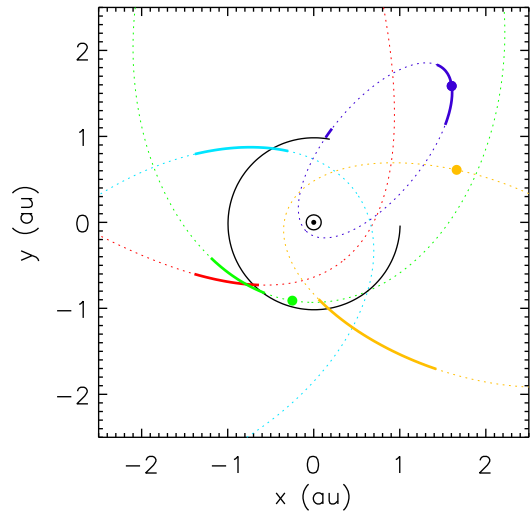}\\
\includegraphics[width=3.in]{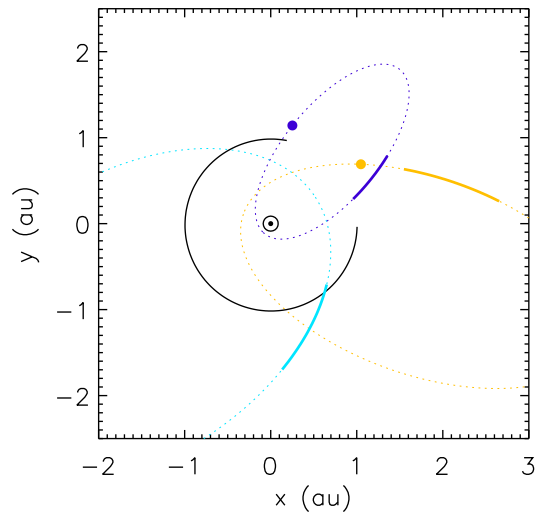} 
\caption{Portions of the orbits where the dust trails are evident are indicated by colored solid lines. 
1P/Halley = red, 2P/Encke = Orange, 73P/SW3 = green, 
169P/NEAT = cyan, 3200 Phaethon = violet. Dotted lines indicate portions of orbits that were not 
observed or where the trails were not evident.
Earth's orbit (black) is only shown for duration of the {\it COBE} cryogenic mission from 89345 - 90264.
The left panel shows sightings of the trails at dates prior to $\sim$90210, the right panel shows after that date.
In each case the trail is only visible for several days when the Earth is near the location where the trail 
crosses the Earth's orbit (or nearest the visible trail segment). Visibility of the trails at small heliocentric
radii is generally limited by DIRBE's elongation limit $\epsilon > 64\arcdeg$. The visibility limits at
large radii are subjective and very uncertain. Locations of the parent bodies (if within the limits of the figure)
are marked as solid dots.}
\label{fig:visibility}
\end{figure}

The portions of the orbits where the dust trails appear to be visible 
are illustrated in Figure \ref{fig:visibility}. DIRBE's elongation 
limit of $\epsilon > 64\arcdeg$ general truncates the minimum heliocentric
radius at which trails can be detected. The limits on the maximum 
heliocentric radius are very subjective
because the trails fade smoothly with respect to time and position 
until they are lost in the confusion of the residual background.
Additionally, at the larger radii, the proper motions of the visible 
trails are smaller and the viewing angles often become nearly tangent to the 
orbit, which means that a large range in radius is mapped into a very small
location on the sky.

\begin{figure}[ht] 
\centering
\includegraphics[width=3.5in]{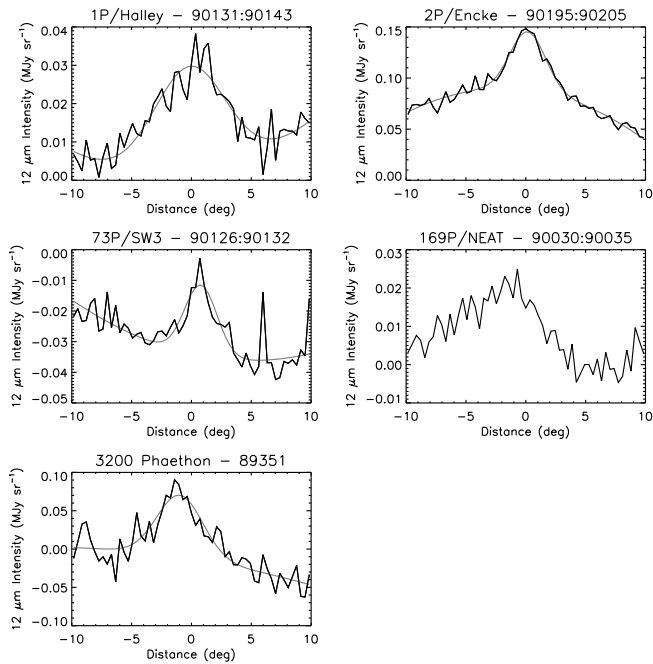}
\caption{Mean 12 $\mu$m profiles of the detected dust trails at the dates indicated.
The gray lines indicated fits using a Gaussian profile and a second-order polynomial background (Eq.~\ref{eq:fit}).
The parameters of the fits are given in Table \ref{tab:profiles}.
\label{fig:profiles_b5}}
\end{figure}

\begin{figure}[ht] 
\centering
\includegraphics[width=3.5in]{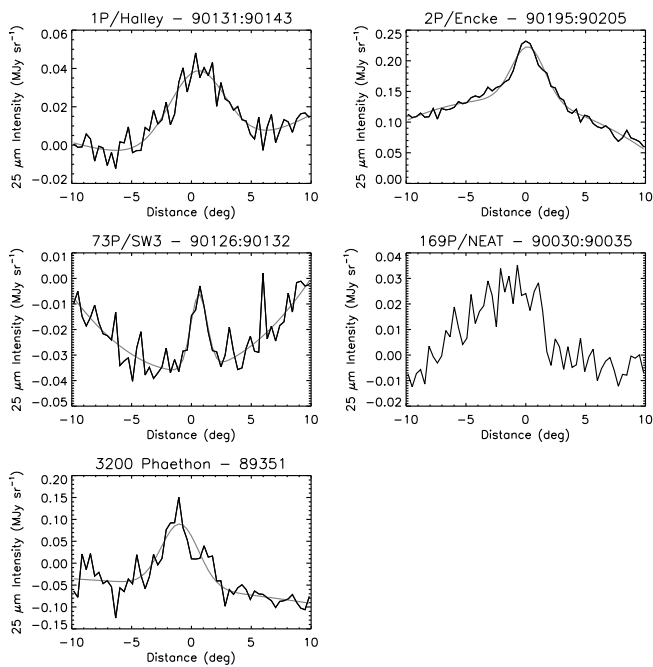} 
\caption{Mean 25 $\mu$m profiles of the detected dust trails, as in Fig. \ref{fig:profiles_b5}.
\label{fig:profiles_b6}}
\end{figure}

Quantifying the brightness and geometry of the comet trails requires 
averaging the emission over the length of the trail (or a fraction thereof)
and as seen over several daily images. To facilitate such measurements,
for each trail the daily images were reprojected into a cartesian 
coordinate system in which the $x$ coordinate is angular distance from the 
perihelion as measured along the orbit, and the $y$ coordinate is the angular
distance perpendicular to the orbit. Mean profiles perpendicular to
the trails were then generated by averaging these images over 
a range of angular distance and a period of time selected for 
good visibility of the trails. 
These intervals are subsets of the 
full range of when and where the trails are visible.
The mean profiles at 12 and 25 $\mu$m 
for each trail are shown in 
Figures \ref{fig:profiles_b5} and \ref{fig:profiles_b6}.

The mean profiles were characterized by fitting them with a Gaussian function 
and a second order polynomial background
\begin{equation}
I_\nu(\theta) = I_0 \exp{[-0.5(\theta-\theta_0)^2/\sigma_\theta^2]} +C_0 + C_1 \theta + C_2 \theta^2.
\end{equation} 
These fits are shown in Figures \ref{fig:profiles_b5} and \ref{fig:profiles_b6} 
and the derived parameters are listed in Table \ref{tab:profiles}. The table lists 
the dates and the length of the trail averaged to generate the profile. 
The approximate ranges in the 
line of sight angle with respect to the orbital plane ($\theta_{LOS}$) and the
mean anomaly of the trail for the length averaged are 
cited.\footnote{Here $\theta_{LOS}$ is defined on the range of $0-90\arcdeg$, 
and can be roughly estimated from $||\beta| - \theta_{LOS}| \leq |i|$ where $\beta$ is the 
observed ecliptic latitude of a point on the orbit and $i$ is the orbital inclination.} 
In detail, these ranges vary during the period integrated.
The profiles for the trails of 2P/Encke and 3200 Phaethon were measured at 
second epochs and are tabulated, but the visibility of the trails at these times 
is poor. The profile of 169P/NEAT was not well fit by this parameterization 
because it is too broadly asymmetrical. 

For all fits, control tests were performed by reflecting each orbit across the ecliptic plane, 
(multiplying the latitude, $\beta$, by $-1$) and then repeating the averaging and fitting
of the data along the reflected orbit. This process samples identical dates, elongations,
and latitudes (in absolute value), but at locations where no trails should be present. 
None of the tests exhibited any indication of a trail, i.e. a resolved gaussian profile above 
the background variations. This confirms that the profiles
shown in Figures \ref{fig:profiles_b5} and \ref{fig:profiles_b6} are very unlikely 
to be random or the result of systematic artifacts.

The peaks of the trail profiles are offset from the orbit of the parent body 
by $\theta_0<1\arcdeg$. The $1\sigma$ dimension of the profiles is
typically $\sigma_\theta \sim1.5\arcdeg$, corresponding to a full width at half maximum 
$FWHM = 2.355\sigma_\theta \approx 5.9\arcdeg$. The trails appear to
be resolved, because the measured width of bright point sources is 
only $\sigma \approx 0.28\arcdeg$ (Gaussian fit). However, some of the apparent 
width of the trails might be caused by the apparent motion of the trail 
(changing parallax) during the course of a day.
 
The mean color of the peak emission of the observed
trails is found to be $I_\nu(25 \micron)/I_\nu(12 \micron) = 1.8\pm0.5$.
The uncertainty listed here is the dispersion between the six successful
fits listed in Table \ref{tab:profiles}. The estimated uncertainty in the 
measurement of each of the flux ratios is similar, assuming $\sim15\%$
uncertainties on the amplitudes of the Gaussian fits.
After applying the appropriate broadband color corrections, the flux ratios 
correspond to a blackbody dust temperature of $T = 281\pm34$~K, 
which is close to the dust temperature of 286~K at 1 au, as
in the \cite{Kelsall:1998} model. This result is to be expected,
given that these trails are primarily detected when closer than $\sim 0.2$~au
(see Tables \ref{tab:nodes} and \ref{tab:crossing}). The colors
of the trails are therefore similar to the of emission other substructures 
in the zodiacal light: the Earth-resonant ring and blobs at 1 au, and 
the migrating asteroidal dust bands (specifically, the inwardly drifting
dust when seen at high latitudes).
 
The peak intensity of the trails is $\lesssim1\%$ of the intensity of the zodiacal light 
at high latitudes. Given the similar dust temperatures, the trails thus have column 
densities that are $\lesssim1\%$ of the zodiacal dust cloud. However the 
observed trails are very limited in extent, and therefore constitute a far smaller fraction
of the total mass of the zodiacal dust.


\begin{deluxetable*}{llllll}
\tabletypesize{\footnotesize}
\tablewidth{0pt}
\tablecaption{DIRBE Comet Trail Profiles}
\tablehead{
\colhead{ } & \colhead{1P/Halley} & \colhead{2P/Encke} & \colhead{73P/SW3} & \colhead{169P/NEAT} & \colhead{3200 Phaethon}
}
\startdata
$q$ (au) & 0.586 & 0.336 & 0.933 & 0.607 & 0.140 \\
$i$ ($\arcdeg$) & 162.26 & 11.78 & 11.42 & 11.31 & 22.18 \\
$P$ (yr) & 75.32 & 3.30 & 5.34 & 4.20 & 1.43 \\
$T_p$ (Julian) & 2446467.4  & 2448193.1  & 2448031.6 & 2447491.0 & 2448196.1 \\
$T_p$ (Yr, Day)  & 86037 & 90302 & 90139 & 88330 & 90305 \\
\hline
Dates Averaged & 90131-90143 & 90195-90205 & 90126-90132 & 90030-90035 & 89351\\
Length Averaged ($\arcdeg$) & 66.4 & 43.9 & 105.1 & 55.5 & 27.8 \\
$R$ Range (au) & 0.97 -- 1.54\tablenotemark{a} & 0.92 -- 1.14 & 1.02 --1.14 & 1.03 -- 1.18  & 1.01 -- 1.05 \\
$\Delta$ Range (au) & 0.14 -- 0.93\tablenotemark{a} & 0.31 -- 0.42 & 0.08 -- 0.32 & 0.24 -- 0.25 & 0.07 -- 0.11\\
$\theta_{LOS}$ Range ($\arcdeg$) & 0 -- 40 & 13 -- 34 & 10 -- 64 & 26 -- 44 & 0 -- 1\\
Mean Anomaly Range ($\arcdeg$) & 0.48 -- 0.97\tablenotemark{a} & 11.2 -- 15.2 & 351.7 -- 354.6 & 10.9 -- 13.6 & 330.9 -- 332.9\\
Comet Mean Anomaly ($\arcdeg$) & 20.4 & 330.4 & 358.1 & 101.7 & 141.9\\
\sidehead{12 $\mu$m Profile}
$I_0$ (MJy sr$^{-1})$ & 0.029 & 0.057 & 0.023 & \nodata & 0.081\\
FWHM $\approx 2.35 \sigma_\theta$ ($\arcdeg$) & 7.1 & 3.5 & 3.3 & \nodata & 4.5\\
$\theta_0$ ($\arcdeg$) & -0.10 & 0.16 & 0.78 & \nodata & -0.96\\
$C_0$ (MJy sr$^{-1})$ & 0.00094 & 0.088 & -0.034 & \nodata & -0.013\\
$C_1$ (MJy sr$^{-1})$ & 0.00040 & -0.0013 & -0.00088 & \nodata & -0.0024\\
$C_2$ (MJy sr$^{-1})$ & 0.00011 & -0.00036 & 8.9e-05 & \nodata & -9.5e-05\\
\sidehead{25 $\mu$m Profile}
$I_0$ (MJy sr$^{-1})$ & 0.042 & 0.089 & 0.030 & \nodata & 0.144\\
FWHM $\approx 2.35 \sigma_\theta$ ($\arcdeg$) & 5.7 & 3.3 & 1.6 & \nodata & 3.8\\
$\theta_0$ ($\arcdeg$) & 0.50 & 0.21 & 0.63 & \nodata & -0.96\\
$C_0$ (MJy sr$^{-1})$ & -0.0039 & 0.13 & -0.036 & \nodata & -0.057\\
$C_1$ (MJy sr$^{-1})$ & 0.00072 & -0.0024 & 0.00033 & \nodata & -0.0028\\
$C_2$ (MJy sr$^{-1})$ & 0.00013 & -0.00056 & 0.00033 & \nodata & -6.3e-05\\
\hline
Dates Averaged & \nodata & 90233-90245 & \nodata & 90208-90218 & 90241-90264\\
Length Averaged ($\arcdeg$) & \nodata & 61.9 & \nodata & 64.3 & 39.7 \\
$R$ Range (au) & \nodata & 1.13 -- 2.43 & \nodata & 1.06 -- 2.09  & 0.93 -- 2.18 \\
$\Delta$ Range (au) & \nodata & 1.16 -- 1.72 & \nodata & 0.17 -- 1.55 & 0.55 -- 1.81\\
$\theta_{LOS}$ Range ($\arcdeg$) &\nodata & 0 -- 1 & \nodata & 4 -- 66 & 10 -- 61\\
Mean Anomaly Range ($\arcdeg$) & \nodata & 311.7 -- 344.9 & \nodata & 327.3 -- 348.5 & 24.0 -- 113.6\\
Comet Mean Anomaly ($\arcdeg$) & \nodata & 342.1 & \nodata & 144.1 & 325.4\\
\sidehead{12 $\mu$m Profile}
$I_0$ (MJy sr$^{-1})$ & \nodata & 0.050 & \nodata & \nodata & 0.016\\
FWHM $\approx 2.35 \sigma_\theta$ ($\arcdeg$) & \nodata & 3.3 & \nodata & \nodata & 3.1\\
$\theta_0$ ($\arcdeg$) & \nodata & -0.83 & \nodata & \nodata & 1.01\\
$C_0$ (MJy sr$^{-1})$ & \nodata & 0.037 & \nodata & \nodata & 0.026\\
$C_1$ (MJy sr$^{-1})$ & \nodata & -0.00074 & \nodata & \nodata & -0.0027\\
$C_2$ (MJy sr$^{-1})$ & \nodata & -7.3e-05 & \nodata & \nodata & 8.7e-05\\
\sidehead{25 $\mu$m Profile}
$I_0$ (MJy sr$^{-1})$ & \nodata & 0.099 & \nodata & \nodata & 0.034\\
FWHM $\approx 2.35 \sigma_\theta$ ($\arcdeg$) & \nodata & 3.5 & \nodata & \nodata & 2.4\\
$\theta_0$ ($\arcdeg$) & \nodata & -0.79 & \nodata & \nodata & 0.93\\
$C_0$ (MJy sr$^{-1})$ & \nodata & 0.0047 & \nodata & \nodata & 0.039\\
$C_1$ (MJy sr$^{-1})$ & \nodata & -0.0031 & \nodata & \nodata & -0.0045\\
$C_2$ (MJy sr$^{-1})$ & \nodata & -0.00021 & \nodata & \nodata & 0.00014
\enddata
\tablecomments{$q$ = perihelion dist., $i$ = inclination, $P$ = period, $T_p$ = Date of perihelion;
{\bf Dates Averaged} = range of dates averaged to generate profile (not necessarily when trail is closest, nor all visible dates),
{\bf Length Averaged} = projected angular length of the trail averaged to generate profile,
{\bf $R$ Range} = range of heliocentric distance within length of trail averaged (varies slightly with date),
{\bf $\Delta$ Range} = range of geocentric distance within length of trail averaged (varies slightly with date),
{\bf $\theta_{LOS}$ Range} = range of angle between LOS and orbital plane (varies with date and position along trail),
{\bf Mean Anomaly Range} = range of mean anomaly measured from perihelion within length of trail averaged (varies slightly with date),
{\bf Comet Mean Anomaly} = Mean anomaly of the parent body measured from perihelion (varies slightly with date);
{\bf profile fit} $ = I_0 \exp{[-0.5(\theta-\theta_0)^2/\sigma_\theta^2]} +C_0 + C_1 \theta + C_2 \theta^2$}
\tablenotetext{a}{Stated ranges are truncated at the tangent point of the orbit.}
\label{tab:profiles}
\end{deluxetable*}

\subsection{Trails and Meteor Showers}

The 5 trails detected by DIRBE make 7 close approaches
to the Earth's orbit during the period of observations. The closest approaches 
occur near the times the orbits cross Earth's orbit (Fig. \ref{fig:orbits})
and/or when the orbits intersects the ecliptic plane.
At 5 of these close approaches 
there are associated meteor showers that have been previously linked
to the parent bodies. These showers are listed in Tables \ref{tab:nodes} and \ref{tab:crossing}.
The only close approach that does not have an associated meteor shower 
(73P/SW3 at day 90212) is relatively distant ($\Delta \gtrsim 0.17$ au)
and no dust trail was visible at the time. 

The DIRBE images were examined with particular attention to 
the parent bodies of the Quadrantid, Perseid, and Leonid meteor
showers. However with the present processing no dust trails 
could be seen. The detection of the associated trails may be 
hampered by less favorable viewing geometries, and, in the case 
of the Leonids, a dust trail that is not fully dispersed along the entire 
orbit of the parent body. 

No dust trail was evident along the orbit
of the Kreutz family comets [e.g. C/1965 S1-A (Ikeya-Seki)]
which initially motivated this study. This is likely because 
the orbits of these comets are highly inclined ($i = 141.8\arcdeg$) to the ecliptic.
Therefore any trail is nearly 1 au distant when viewed toward the SEP, and
never closer than $\sim0.6$ au at lower solar elongations. These 
minimum distances are far greater than those of any dust trails 
detected by DIRBE so far.

In general the trails that are detected are found at large distances (mean anomalies) 
from their parent bodies, and thus one might regard them as meteoroid streams 
rather than more traditional dust trails. An exception is the dust trail observed for 
73P/SW3, which is only seen in relatively close association with (and trailing) the parent comet.
It is interesting to note that these DIRBE observations show that 73P/SW3 had 
a prominent dust trail prior to its breakup in 1995 at its next perihelion passage \citep{crovisier:1996}.
The trail observed by \cite{Reach:2009} with {\it Spitzer} in 2006 is $\sim6$ 
times brighter than the DIRBE measurement at 24 $\mu$m, although 
this comparison may be strongly influenced by the large difference in angular resolution.
{\it Spitzer} observations of the dust trail of 2P/Encke \citep{Reach:2007} are similarly 
bright compared to the DIRBE measurements reported here, though the DIRBE observation 
are much more distant from the comet.

The portions of the trail of asteroid 3200 Phaethon that are seen moving near the ecliptic
poles are also relatively far from the parent body. However, when crossing the 
line of nodes near Day 89351, the trail along a distant portion of the orbit appears 
to be visible, especially at 25 $\mu$m. Phaethon is embedded 
within this segment (Fig. \ref{fig:visibility}a), although too faint to detect with DIRBE. 
Recent reports of ongoing dust ejection from 
Phaethon \citep{Li:2013,Jewitt:2013} have noted strong activity for short intervals 
(a few days) immediately after perihelion. Since Phaethon is nearer aphelion 
during the DIRBE observations, the trail segments seen here are likely only 
constraints on past (though perhaps recent) dust production, rather than evidence of 
active ongoing dust production.

\subsection{Future Prospects} 
  This paper is only an introduction to the possibilities of using the DIRBE data 
for the study of comet dust trails. The techniques presented above are sufficient to 
find the brightest dust trails, but there are several lines of investigation that 
may lead to more accurate and sensitive measurement of these and other trails.
For example: \\
(1) Ideally one would perform this analysis after subtraction of a perfect model
of the emission from the main interplanetary dust cloud of cometary and asteroidal 
dust. Alternate models of the zodiacal light that might yield improved results have
been presented by (e.g.) \cite{Wright:1998} and \cite{Rowan-Robinson:2013}.\\
(2) Lacking a perfect zodiacal light model, some additional ad hoc removal of residual 
emission is still likely to improve the visibility of the comet trails. There are many other ways that 
the residual images could be filtered or processed to remove residual zodiacal light
and instrumental effects. However, any such processing schemes must be careful to 
avoid removing the emission of the trails along with the unwanted artifacts. It may 
be that such processing needs to be altered on a case by case basis, optimized for
each particular trail.\\
(3) At the shorter wavelengths ($\leq 4.9$ $\mu$m), residual artifacts at the edges of bright 
point sources are a major limitation to recognizing low surface brightness structures.
This issue could be attacked with modified map-making procedures. Super-resolved
images may allow more detailed and accurate mapping of each point source, though 
this  would dilute the effective coverage (depth) of the images. Conversely, the images 
could be mapped at (or convolved to) sufficiently low resolution, such that details of the 
beam shape are irrelevant to the reconstruction of the images.\\
(4) Alternately, one may forego map-making altogether and extract information 
on the dust trails directly from the time domain data. An advantage to this approach
is that it would avoid the daily averaging of the trails, which may artificially broaden and 
weaken the trails, especially in cases where their proper motion is high.
The disadvantage here is that in the time domain it may be difficult to find 
depictions of the data that clearly show the trails, or that could be used to 
search for additional trails.\\
(5) More focussed attention could be paid to non-Earth-crossing comets and trails.
These trails would general appear (approximately) as great circles, with low inclinations.
Throughout the year, they would always appear as bands at low to moderate ecliptic
latitudes, where they could easily be confused with the brighter asteroidal 
dust bands.

The DIRBE results presented here suggest that dust trails may be more
prevalent than previously expected.  Searches for trails in archival data sets 
should not be limited to looking near the parent bodies, but should also 
focus on (a) times when the trails are especially close to the Earth's orbit,
and projected at high ecliptic latitudes in the case of Earth-crossing objects,
and (b) the possibility of detecting structures that may be $>1\arcdeg$ in width.
In all cases, the data reduction must take care that the signal from 
very extended, low surface brightness, {\it and moving} emission is
not accidentally removed.

\section{Summary} 
The DIRBE data have been reprocessed for the purpose of looking 
for comet dust trails. The current procedure creates average 
images on a daily basis rather than a weekly basis. These images 
have zodiacal light subtracted according to the \cite{Kelsall:1998} model, 
and have an additional subtraction of the slow temporal variation (12 and 6
month periods) of the residual emission and fixed background.
Animations of these daily images are effective 
for identifying moving sources within the solar system, including faint objects
(asteroids and comets) and low surface brightness structures (dust trails) that are difficult
to identify in a single image. One comet and 13 asteroids were found, 
in addition to the 4 comets and 3 asteroids that had been previously noted.
Five new and existing comet dust trails are observed by DIRBE, each
 associated with Earth-crossing 
objects, which have perihelions $q < 1$ au. 
The trails are most clearly seen when closest to the Earth orbit, and
at moderate to high ecliptic latitude. Some of these trails can be seen far 
from their parent comets (or asteroid). All the trails are associated with parent bodies
of established meteor showers, although not all major meteor showers have
evident IR emission. Further work on DIRBE data should 
be able to extend wavelength coverage, and may be able to reveal additional fainter trails.

\acknowledgments
We thank Tom Kelsall for inspiring the methods used to 
remove residual variations of the zodiacal light, and Rich Barry for 
working out prototype tools used to read and process the DIRBE CIO
data set. The referee, Mark Sykes, is thanked for useful discussion
which resulted in a clearer, more convincing presentation and additional 
characterization of the trails.
This work was partially supported by NASA ROSES-ADAP grant NNX08AW22G.
The analysis made extensive use of IDL programs from the IDLASTRO
library\footnote{\url{http://idlastro.gsfc.nasa.gov}} and from the 
library of Marc Buie\footnote{\url{http://www.boulder.swri.edu/~buie/idl/}}.
This research has made use of NASA's Astrophysics Data System.
We acknowledge the use of the Legacy Archive for Microwave Background Data 
Analysis (LAMBDA), part of the High Energy Astrophysics Science Archive Center 
(HEASARC). HEASARC/LAMBDA is a service of the Astrophysics Science Division at 
the NASA Goddard Space Flight Center.\\

{\it Facility:} \facility{COBE (DIRBE)}.

\bibliographystyle{apj}
\bibliography{trails}

\begin{thebibliography}{}
\expandafter\ifx\csname natexlab\endcsname\relax\def\natexlab#1{#1}\fi

\bibitem[{{Boggess} {et~al.}(1992){Boggess}, {Mather}, {Weiss}, {Bennett},
  {Cheng}, {Dwek}, {Gulkis}, {Hauser}, {Janssen}, {Kelsall}, {Meyer},
  {Moseley}, {Murdock}, {Shafer}, {Silverberg}, {Smoot}, {Wilkinson}, \&
  {Wright}}]{Boggess:1992}
{Boggess}, N.~W., {Mather}, J.~C., {Weiss}, R., {et~al.} 1992, \apj, 397, 420

\bibitem[{{Burns} {et~al.}(1979){Burns}, {Lamy}, \& {Soter}}]{Burns:1979}
{Burns}, J.~A., {Lamy}, P.~L., \& {Soter}, S. 1979, Icarus, 40, 1

\bibitem[{{Crovisier} {et~al.}(1996){Crovisier}, {Bockelee-Morvan}, {Gerard},
  {Rauer}, {Biver}, {Colom}, \& {Jorda}}]{crovisier:1996}
{Crovisier}, J., {Bockelee-Morvan}, D., {Gerard}, E., {et~al.} 1996, \aap, 310,
  L17

\bibitem[{{Giorgini} {et~al.}(1996){Giorgini}, {Yeomans}, {Chamberlin},
  {Chodas}, {Jacobson}, {Keesey}, {Lieske}, {Ostro}, {Standish}, \&
  {Wimberly}}]{Giorgini:1996}
{Giorgini}, J.~D., {Yeomans}, D.~K., {Chamberlin}, A.~B., {et~al.} 1996, in
  Bulletin of the American Astronomical Society, Vol.~28, AAS/Division for
  Planetary Sciences Meeting Abstracts \#28, 1158

\bibitem[{{Greisen} {et~al.}(2006){Greisen}, {Calabretta}, {Valdes}, \&
  {Allen}}]{Greisen:2006}
{Greisen}, E.~W., {Calabretta}, M.~R., {Valdes}, F.~G., \& {Allen}, S.~L. 2006,
  \aap, 446, 747

\bibitem[{{Hauser} {et~al.}(1998){Hauser}, {Arendt}, {Kelsall}, {Dwek},
  {Odegard}, {Weiland}, {Freudenreich}, {Reach}, {Silverberg}, {Moseley},
  {Pei}, {Lubin}, {Mather}, {Shafer}, {Smoot}, {Weiss}, {Wilkinson}, \&
  {Wright}}]{Hauser:1998}
{Hauser}, M.~G., {Arendt}, R.~G., {Kelsall}, T., {et~al.} 1998, \apj, 508, 25

\bibitem[{{Ipatov} {et~al.}(2008){Ipatov}, {Kutyrev}, {Madsen}, {Mather},
  {Moseley}, \& {Reynolds}}]{Ipatov:2008}
{Ipatov}, S.~I., {Kutyrev}, A.~S., {Madsen}, G.~J., {et~al.} 2008, Icarus,
  194, 769

\bibitem[{{Ishiguro} {et~al.}(2009){Ishiguro}, {Sarugaku}, {Nishihara},
  {Nakada}, {Nishiura}, {Soyano}, {Tarusawa}, {Mukai}, {Kwon}, {Hasegawa},
  {Usui}, \& {Ueno}}]{Ishiguro:2009}
{Ishiguro}, M., {Sarugaku}, Y., {Nishihara}, S., {et~al.} 2009, Advances in
  Space Research, 43, 875

\bibitem[{{Jenniskens}(2006)}]{Jenniskens:2006}
{Jenniskens}, P. 2006, {Meteor Showers and their Parent Comets} (Cambridge:
  Cambridge Universty Press)

\bibitem[{{Jewitt} {et~al.}(2013){Jewitt}, {Li}, \& {Agarwal}}]{Jewitt:2013}
{Jewitt}, D., {Li}, J., \& {Agarwal}, J. 2013, \apjl, 771, L36

\bibitem[{{Kelsall} {et~al.}(1998){Kelsall}, {Weiland}, {Franz}, {Reach},
  {Arendt}, {Dwek}, {Freudenreich}, {Hauser}, {Moseley}, {Odegard},
  {Silverberg}, \& {Wright}}]{Kelsall:1998}
{Kelsall}, T., {Weiland}, J.~L., {Franz}, B.~A., {et~al.} 1998, \apj, 508, 44

\bibitem[{Kirkwood(1873)}]{Kirkwood:1873}
Kirkwood, D. 1873, Comets and Meteors: Their Phenomena in All Ages; Their
  Mutual Relations; and the Theory Of Their Origin (Philadelphia: J.B.
  Lippincott \& Co.)

\bibitem[{{Knight} {et~al.}(2010){Knight}, {A'Hearn}, {Biesecker}, {Faury},
  {Hamilton}, {Lamy}, \& {Llebaria}}]{Knight:2010}
{Knight}, M.~M., {A'Hearn}, M.~F., {Biesecker}, D.~A., {et~al.} 2010, \aj, 139,
  926

\bibitem[{{Knight} \& {Battams}(2014)}]{Knight:2014}
{Knight}, M.~M., \& {Battams}, K. 2014, \apjl, 782, L37

\bibitem[{{Li} \& {Jewitt}(2013)}]{Li:2013}
{Li}, J., \& {Jewitt}, D. 2013, \aj, 145, 154

\bibitem[{{Lisse} {et~al.}(1998){Lisse}, {A'Hearn}, {Hauser}, {Kelsall},
  {Lien}, {Moseley}, {Reach}, \& {Silverberg}}]{Lisse:1998}
{Lisse}, C.~M., {A'Hearn}, M.~F., {Hauser}, M.~G., {et~al.} 1998, \apj, 496,
  971

\bibitem[{{Littmann}(1998)}]{Littmann:1998}
{Littmann}, M. 1998, {The Heavens on Fire} (Cambridge: Cambridge University
  Press)

\bibitem[{{Low} {et~al.}(1984){Low}, {Young}, {Beintema}, {Gautier},
  {Beichman}, {Aumann}, {Gillett}, {Neugebauer}, {Boggess}, \&
  {Emerson}}]{Low:1984}
{Low}, F.~J., {Young}, E., {Beintema}, D.~A., {et~al.} 1984, \apjl, 278, L19

\bibitem[{{Marsden}(1989)}]{Marsden:1989}
{Marsden}, B.~G. 1989, \aj, 98, 2306

\bibitem[{{Marsden}(2005)}]{Marsden:2005}
---. 2005, \araa, 43, 75

\bibitem[{{Nesvorn{\'y}} {et~al.}(2006){Nesvorn{\'y}}, {Sykes}, {Lien},
  {Stansberry}, {Reach}, {Vokrouhlick{\'y}}, {Bottke}, {Durda}, {Jayaraman}, \&
  {Walker}}]{Nesvorny:2006}
{Nesvorn{\'y}}, D., {Sykes}, M., {Lien}, D.~J., {et~al.} 2006, \aj, 132, 582

\bibitem[{{Neugebauer} {et~al.}(1984){Neugebauer}, {Habing}, {van Duinen},
  {Aumann}, {Baud}, {Beichman}, {Beintema}, {Boggess}, {Clegg}, {de Jong},
  {Emerson}, {Gautier}, {Gillett}, {Harris}, {Hauser}, {Houck}, {Jennings},
  {Low}, {Marsden}, {Miley}, {Olnon}, {Pottasch}, {Raimond}, {Rowan-Robinson},
  {Soifer}, {Walker}, {Wesselius}, \& {Young}}]{Neugebauer:1984}
{Neugebauer}, G., {Habing}, H.~J., {van Duinen}, R., {et~al.} 1984, \apjl, 278,
  L1

\bibitem[{{Reach} {et~al.}(2007){Reach}, {Kelley}, \& {Sykes}}]{Reach:2007}
{Reach}, W.~T., {Kelley}, M.~S., \& {Sykes}, M.~V. 2007, Icarus, 191, 298

\bibitem[{{Reach} {et~al.}(2009){Reach}, {Vaubaillon}, {Kelley}, {Lisse}, \&
  {Sykes}}]{Reach:2009}
{Reach}, W.~T., {Vaubaillon}, J., {Kelley}, M.~S., {Lisse}, C.~M., \& {Sykes},
  M.~V. 2009, Icarus, 203, 571

\bibitem[{{Rowan-Robinson} \& {May}(2013)}]{Rowan-Robinson:2013}
{Rowan-Robinson}, M., \& {May}, B. 2013, \mnras, 429, 2894

\bibitem[{{Schiaparelli}(1867)}]{Schiaparelli:1867}
{Schiaparelli}, M.~J.~V. 1867, Astronomische Nachrichten, 68, 331

\bibitem[{{Silverberg} {et~al.}(1993){Silverberg}, {Hauser}, {Boggess},
  {Kelsall}, {Moseley}, \& {Murdock}}]{Silverberg:1993}
{Silverberg}, R.~F., {Hauser}, M.~G., {Boggess}, N.~W., {et~al.} 1993, in
  Society of Photo-Optical Instrumentation Engineers (SPIE) Conference Series,
  Vol. 2019, Infrared Spaceborne Remote Sensing, ed. M.~S. {Scholl}, 180--189

\bibitem[{{Smith} {et~al.}(2004){Smith}, {Price}, \& {Baker}}]{Smith:2004}
{Smith}, B.~J., {Price}, S.~D., \& {Baker}, R.~I. 2004, \apjs, 154, 673

\bibitem[{{Sykes}(1988)}]{Sykes:1988}
{Sykes}, M.~V. 1988, \apjl, 334, L55

\bibitem[{{Sykes} {et~al.}(1986){Sykes}, {Lebofsky}, {Hunten}, \&
  {Low}}]{Sykes:1986}
{Sykes}, M.~V., {Lebofsky}, L.~A., {Hunten}, D.~M., \& {Low}, F. 1986, Science,
  232, 1115

\bibitem[{{Sykes} {et~al.}(1990){Sykes}, {Lien}, \& {Walker}}]{Sykes:1990}
{Sykes}, M.~V., {Lien}, D.~J., \& {Walker}, R.~G. 1990, Icarus, 86, 236

\bibitem[{{Sykes} \& {Walker}(1992)}]{Sykes:1992}
{Sykes}, M.~V., \& {Walker}, R.~G. 1992, Icarus, 95, 180

\bibitem[{{Vaubaillon} \& {Reach}(2010)}]{Vaubaillon:2010}
{Vaubaillon}, J.~J., \& {Reach}, W.~T. 2010, \aj, 139, 1491

\bibitem[{{Wright}(1998)}]{Wright:1998}
{Wright}, E.~L. 1998, \apj, 496, 1

\end{thebibliography}

\end{document}